\documentclass[twocolumn]{jpsj3} 
\usepackage{bm}
\usepackage{mathrsfs}
\usepackage{exscale}
\usepackage{color}

\def\runauthor{K.~Deguchi~\textit{et al.}}

\title{Absence of Meissner State and Robust Ferromagnetism in the Superconducting State of UCoGe:
Possible Evidence of Spontaneous Vortex State}

\author{
Kazuhiko~\textsc{Deguchi}$^{}$\thanks{E-mail address: deguchi@edu3.phys.nagoya-u.ac.jp},
Eisuke~\textsc{Osaki}$^{}$,
Seiko~\textsc{Ban}$^{}$,
Nobuyuki~\textsc{Tamura}$^{}$,
Yasuyuki~\textsc{Simura}$^{1}$,
Toshiro~\textsc{Sakakibara}$^{1}$,
Isamu~\textsc{Satoh}$^{2}$,
and
Noriaki~K.~\textsc{Sato}$^{}$
}

\inst{
$^{}$Department of Physics, Graduate School of Science, Nagoya University, Nagoya 464-8602, Japan\\
$^{1}$Institute for Solid State Physics, The University of Tokyo, Kashiwa 277-8581, Japan\\
$^{2}$Institute for Materials Research, Tohoku University, Sendai 980-8577 Japan
}

\abst{
We report ac magnetic susceptibility and dc magnetization measurements on the superconducting ferromagnet UCoGe (with superconducting and Curie temperatures of $T_{{\rm SC}} \sim 0.5$~K and $T_{{\rm Curie}} \sim 2.5$~K, respectively).
In the normal, ferromagnetic state ($T_{{\rm SC}} < T < T_{{\rm Curie}}$), the magnetization curve exhibits a hysteresis loop similar to that of a regular itinerant ferromagnet. 
Upon lowering the temperature below $T_{{\rm SC}}$, the spontaneous magnetization is unchanged, but the hysteresis is markedly enhanced. 
Even deeply inside the superconducting state, ferromagnetism is not completely shielded, and there is no Meissner region, a magnetic field region of $H < H_{\rm c1}$ (a lower critical field).
From these results, we suggest that UCoGe is the first material in which ferromagnetism robustly survives in the superconducting state and a spontaneous vortex state without the Meissner state is realized.  
}

\kword{superconducting ferromagnet, unconventional superconductivity, U-based heavy fermion, UCoGe, magnetization, spontaneous vortex state} 

\begin {document}
\maketitle

The interplay of superconductivity and ferromagnetism, which would be antagonistic because of the competitive nature between the screening by the Meissner effect and internal fields generated by magnetic orderings, has been of great interest~\cite{Ginzburg1957}. 
The following materials were considered as promising candidates for their coexistence; ErRh$_4$B$_4$, HoMo$_6$S$_8$~\cite{Bulaevskii1985}, ErNi$_2$B$_2$C~\cite{Canfield1996}, RuSr$_2$RE$_2$Cu$_2$O$_{10-\delta}$~\cite{Felner1997}, and RuSr$_2$RECu$_2$O$_8$~\cite{Bernhard1999}. 
From the results of extensive studies, however, it was found that ferromagnetism suffers antiferromagnetic modifications in the superconducting state~\cite{Bulaevskii1985}. 
Theoretically, various types of modification, such as those on cryptoferromagnetism~\cite{Anderson1959}, spiral structure~\cite{Blount1979}, and domain structure~\cite{Krey1972}, are known. 
In some cases, there is no long range order.
Therefore, the ``ferromagnetism'' coexisting with superconductivity is not real ferromagnetism in these materials.

A new state, known as a spontaneous vortex state or a self-induced vortex state, was proposed to emerge as a compromise of the two competitive orders~\cite{Tachiki1980,Kuper1980,Greenside1981}. 
This differs from a mixed state, i.e., a state of $H_{\rm c1} < H < H_{\rm c2}$ (an upper critical field), of a regular type-II superconductor in that the spontaneous magnetization of ferromagnetism plays a role analogous to external fields, and hence, there is no Meissner state; here and throughout the paper, the Meissner state means a region of $H < H_{\rm c1}$, where all magnetic fluxes are expelled from samples. 
The criterion for the spontaneous vortex state is therefore the presence/absence of a lower critical field $H_{\rm c1}$ in the deep $(T \rightarrow 0)$ superconducting state.

Recent observations of superconductivity in uranium-based superconducting magnets, UGe$_2$~\cite{Saxena2000}, URhGe~\cite{Aoki2001}, and UCoGe~\cite{Huy2007} have renewed our interest in the interplay.
Questions may arise whether the Meissner state is present and whether the ferromagnetism survives in their superconducting state.
In this respect, UCoGe can be a good experimental platform to study the questions, because the superconductivity appears at ambient pressure.

UCoGe is a ferromagnet characterized by a low Curie temperature $T_{{\rm Curie}} \sim 2.5$~K and a small ordered moment $M_0 \sim 0.03~\mu_{\rm B}$/U. 
The ratio of the Curie-Weiss effective moment $p_{\rm eff} \sim 1.7\mu_{\rm B}$ to $M_0$ is very large ($p_{\rm eff}/M_0 \sim 57$),~\cite{Huy2007} and hence, UCoGe is classified as a weak itinerant ferromagnet on the Rhodes-Wohlfarth plot~\cite{Rhodes1963}.
The ising-type anisotropic behavior along the orthorhombic $c$-axis was observed in magnetization measurements for $T \geq 2$~K~\cite{Huy2008}.
Interestingly, superconductivity appears below $T_{{\rm SC}} \sim 0.5$~K.
It might be expected that itinerant $5f$ electrons of uranium atoms contribute to both ferromagnetism and superconductivity.
Nuclear quadrupole resonance (NQR)~\cite{Ohta2010} and muon spin rotation ($\mu$SR)~\cite{Visser2009} experiments revealed the microscopic coexistence of the weak ferromagnetic order with superconductivity. 
Measurements of the nuclear spin-lattice relaxation rate $1/T_1$ in Co-NQR~\cite{Ohta2008} and the upper critical field $H_{c2}$~\cite{Huy2008,Aoki2009} provide evidence for unconventional superconductivity with spin-triplet pairs. 
In this Letter, we report ac susceptibility and dc magnetization measurements in the superconducting state of UCoGe, and show that the Meissner state is absent and ferromagnetism remains unchanged in the superconducting state.
These results establish the intrinsic coexistence of superconductivity and ferromagnetism, which will shed new light on the long-standing issue.

A single crystal of UCoGe was grown by the Czochralski pulling method in a tetra-arc furnace under a high-purity argon atmosphere.
Samples for the ac susceptibility and dc magnetization measurements were cut by spark erosion from the single crystalline ingot into two small pieces: sample \#2 with dimensions of $1.65 \times 1.65 \times 1.89$~mm$^3$ and a mass of 55.8~mg, and sample \#3 with dimensions of $1.50 \times 1.50 \times 3.10$~mm$^3$ and a mass of 83.4~mg.  
These samples were not heat treated. Nevertheless, they exhibited a large residual resistivity ratio ($RRR$) of approximately 20 along the orthorhombic $a$-axis. 
They also exhibited a jump of the specific heat, confirming the bulk superconductivity.
Note that sample \#2 was used for the NQR experiment which showed that the entire volume of the sample becomes ferromagnetic and superconductivity microscopically coexists with ferromagnetism~\cite{Ohta2010}, indicating that the superconducting and ferromagnetic regions are not separately present.

For the magnetization measurement at low temperatures down to about 0.1~K, we used a capacitive Faraday force magnetometer installed in a $^3$He-$^4$He dilution refrigerator~\cite{Sakakibara1994}. 
The resolution of the system is higher than 10$^{-5}$~emu. 
Throughout the measurement, we applied a field gradient (0.1~kOe/cm), which is indispensable to this method. 
The resulting field distribution inside the sample was estimated to be less than 30~Oe. 
The magnetic field $H$ was applied along the easy $c$-axis by a superconducting magnet.
The ac susceptibility was measured using a driving field with an excitation frequency of 100~Hz and a magnitude of $H_{\rm ac} \sim 0.1$~Oe, where $H_{\rm ac} \parallel H$.

\begin{figure}[t]
\begin{center}
\includegraphics[clip,width=1.00\columnwidth]{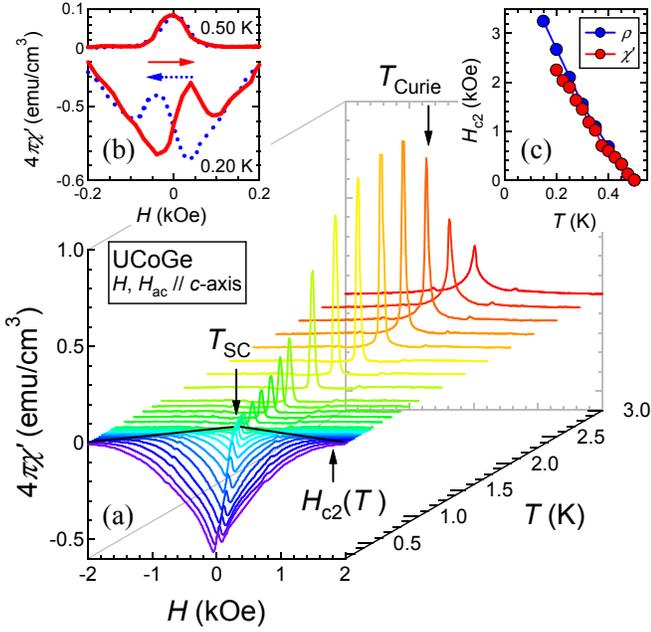}
\end{center}
\caption{(Color online) (a) Field ($H$)-temperature ($T$) dependence of the ac magnetic susceptibility $\chi'(H,T)$ of UCoGe for $H \parallel c$-axis ($-2 \leq H \leq 2$~kOe) at temperatures from $0.2$ to $3.0$~K. (For clarity, we show only increasing field data here.) Arrows at zero field indicate anomalies associated with the following two phase transitions: superconductivity at $T_{\rm SC} = 0.48$~K and ferromagnetism at $T_{\rm Curie} = 2.55$~K. Black lines denote the upper critical field $H_{\rm c2}(T)$. (b) Magnetic field dependence of the ac susceptibility of UCoGe at small fields and $T = 0.5$~K ($> T_{\rm SC}$) and $T = 0.2$~K ($< T_{\rm SC}$). Solid and dotted lines denote increasing- and decreasing-field ac susceptibility, respectively. (c) Temperature dependence of $H_{\rm c2}(T)$ deduced from the ac susceptibility ($\chi'$) and resistivity ($\rho$) measurements; see text for detailed definition.}
\label{fig:Chi-H}
\end{figure}
\begin{figure}[t]
\begin{center}
\includegraphics[clip,width=1.00\columnwidth]{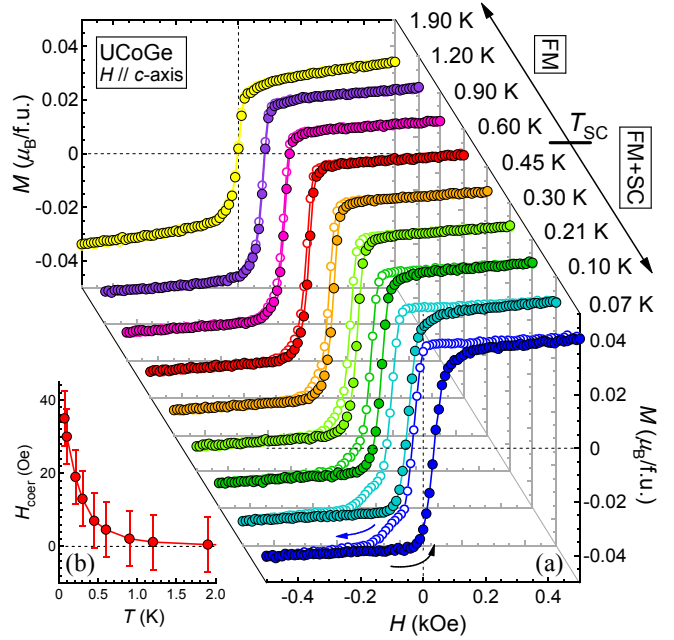}
\end{center}
\caption{(Color online) (a) Temperature variation of the magnetization curves of UCoGe in magnetic field $H \parallel c$-axis ($-0.5 \leq H \leq 0.5$~kOe) at temperatures from $0.07$ to $1.9$~K. Closed (open) symbols denote increasing- and decreasing-field magnetizations. (b) Temperature dependence of the coercive field $H_{\rm coer}$ (half-width of the hysteresis loop) as a function of temperature.}
\label{fig:M-H}
\end{figure}

Figure~\ref{fig:Chi-H}(a) shows the real part of the ac magnetic susceptibility $\chi'(H,T)$ of sample \#2 as functions of external magnetic field $H$ and temperature $T$.
A sharp peak at $H \simeq 0$ grows with lowering temperature until reaching a maximum around $T_{\rm Curie} = 2.55$~K. 
At this temperature, an anomaly appears in the electrical resistivity and the specific heat (not shown here). 
When further cooling the sample, a sharp drop of $\chi'(T)$ occurs at $T_{\rm SC} = 0.48$~K owing to the transition into the superconducting state.
This value of $T_{\rm SC}$ almost equals the transition temperature deduced from zero resistivity, but it is lower than the resistive onset temperature of 0.65~K.

Figure~\ref{fig:Chi-H}(b) shows the $H$ dependence of $\chi'(H,T)$ at $T = 0.5$~K ($ > T_{\rm SC}$) and $T = 0.2$~K ($ < T_{\rm SC}$).
The sharp peak around zero field at $T = 0.5$~K is a part of the butterfly-shaped hysteresis loop, which is characteristic of ferromagnetism\cite{Salas1993}. 
Interestingly, $\chi'(H,T)$ at $T= 0.2$~K differs markedly from that of a regular superconductor that shows a complete shielding at low fields below $H_{\rm c1}$. 
We should note two points in the ac susceptibility measurement that is quite sensitive to magnetic shielding: First, the ferromagnetism is not completely shielded and manifests itself even in zero external field deeply inside the superconducting state. 
Second, a superconducting diamagnetic signal appears to be superposed on the ferromagnetic signal and decreases monotonically with increasing $H$. 
Both of these results strongly suggest the absence of $H_{\rm c1}$.

Figure~\ref{fig:Chi-H}(c) shows $H_{\rm c2}(T)$ determined by the following two definitions: an onset of diamagnetism (denoted $\chi'$) and zero resistivity (denoted $\rho$). 
Both definitions provide good agreement.
We observe the $H_{\rm c2}(T)$ curves to exhibit an unusual upturn behavior with lowering temperature, which is consistent with reported results~\cite{Huy2008,Aoki2009}.

Figure~\ref{fig:M-H}(a) shows the $H$-$T$ dependence of the magnetization $M(H,T)$ of sample \#3 in the temperature range between $0.07$ and $1.9$~K.  Below $T_{\rm Curie} = 2.55$~K, a spontaneous magnetization of $M_{0} \sim 0.04~\mu_{\rm B}$/U is clearly observed. 
As in a regular ferromagnet, $M(H)$ reverses its sign (i.e., $\vec M(H)$ reverses its direction) in a narrow field interval, resulting in the formation of a hysteresis loop, albeit with a small coercive field $H_{\rm coer}$. 
A large high-field susceptibility (i.e., non-saturating behavior of the magnetization at high fields) confirms the weak itinerant ferromagnetism of UCoGe. 
It is noteworthy that the overall feature characteristic of ferromagnetism is retained in the superconducting state.
(For the steep increase in $H_{\rm coer}$ below $T_{\rm SC}$ (Fig.~\ref{fig:M-H}(b)), we discuss its origin below.)
From this result, we confirm that ferromagnetism survives in the superconducting state. This is consistent with the observation of the butterfly-shaped hysteresis loop in the ac susceptibility measurement.

\begin{figure}[t]
\begin{center}
\includegraphics[clip,width=1.00\columnwidth]{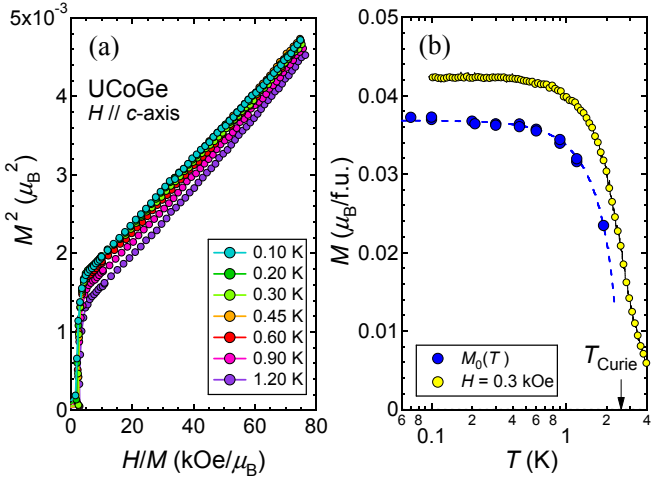}
\end{center}
\caption{(Color online) (a) Arrott plots of magnetization isotherms measured in magnetic field $H \parallel c$-axis ($-5 \leq H \leq 5$~kOe) at temperatures from $0.1$ to $1.2$~K. For the measurement, a field gradient ($0.5$~kOe/cm) was superposed on the dc field $H$. (b) Temperature dependence of $M_{0}(T)$ and $M(T)$. Here, $M_{0}(T)$ is the spontaneous magnetization determined from an extrapolation to zero field in the Arrott plots, and $M(T)$ the magnetization measured at $H = 0.3$~kOe parallel to the $c$-axis under field-cooled condition.}
\label{fig:Arrott}
\end{figure}
\begin{figure}[t]
\begin{center}
\includegraphics[clip,width=1.00\columnwidth]{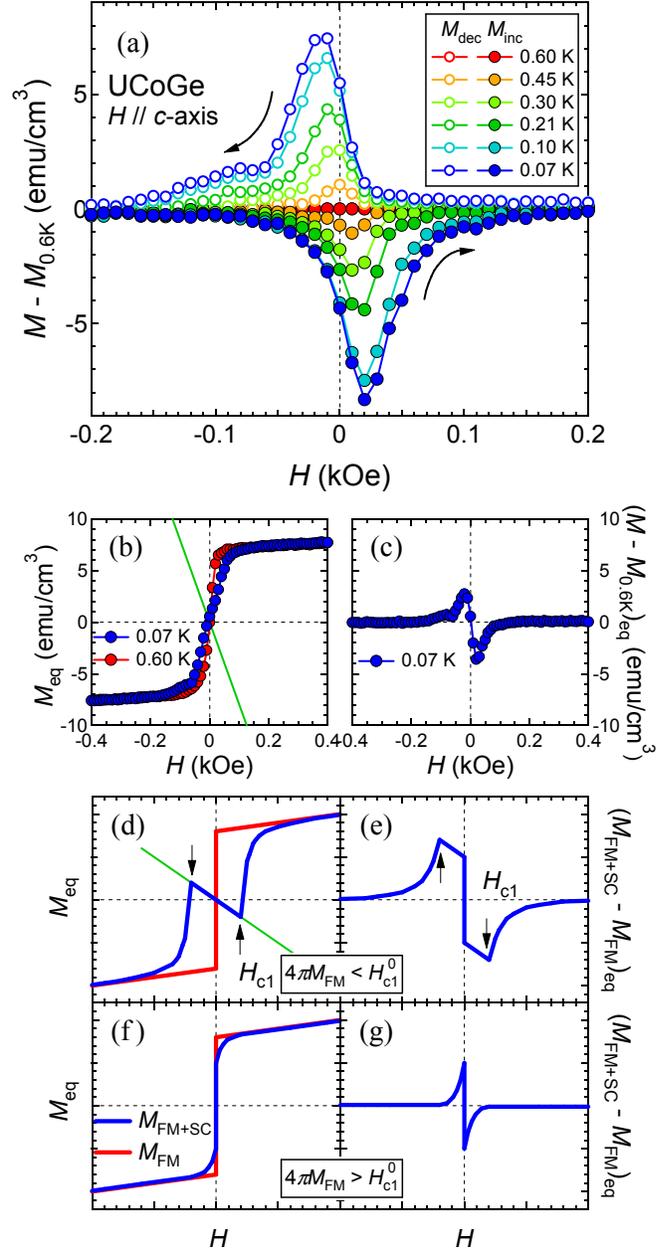}
\end{center}
\caption{(Color online) (a) Field variation of the magnetization difference $M(H,T)-M(H,0.6{\rm K})$ at temperatures from $0.07$ to $0.6$~K of UCoCe. Closed (open) symbols denote increasing- and decreasing-$H$ magnetizations. (b) Field variation of the equilibrium magnetization $M_{\rm eq}(H,T) = [M_{\rm inc}(H,T)+M_{\rm dec}(H,T)]/2$ at $T = 0.07$  and $0.6$~K, where $M_{\rm inc}(H,T)$ and $M_{\rm dec}(H,T)$ are increasing- and decreasing-$H$ magnetization, respectively. The green line denotes $M=-1/4\pi H$, which is expected from the Meissner effect. (c) Magnetization difference $[M(H,T)-M(H,0.6{\rm K})]_{\rm eq}$ at $T = 0.07$~K. (d)(e) Schematic diagrams of $M_{\rm eq}(H)$ and the magnetization difference $[M_{\rm FM+SC}(H)-M_{\rm FM}(H)]_{\rm eq}$ for the case of $4\pi M_{\rm FM} < H_{\rm c1}^{0}$. See text for the definitions of $M_{\rm FM+SC}$, $M_{\rm FM}$, and $H_{\rm c1}^{0}$. (f)(g) $M_{\rm eq}(H)$ and $[M_{\rm FM+SC}(H)-M_{\rm FM}(H)]_{\rm eq}$ for the case of $H_{\rm c1}^{0} < 4\pi M_{\rm FM}$. }
\label{fig:equilibrium}
\end{figure}

Let us examine the detailed $T$ dependence of the spontaneous magnetization $M_{0}(T)$ above and below $T_{\rm SC}$.
Figure~\ref{fig:Arrott}(a) shows Arrott plots of magnetization curves for $-5 \leq H \leq 5$~kOe at temperatures from $0.1$ to $1.2$~K. 
We find good linearity between $M^2$ and $H/M$, indicating that the Arrott plots are useful for the analysis of this system. 
Then, we obtain $M_{0}(T)$ by linear extrapolation to zero external magnetic field.  
The result is plotted in Fig.~\ref{fig:Arrott}(b) along with $M(T)$ measured at $H = 0.3$~kOe.
We find no discernible change in the vicinity of $T_{\rm SC}$.
This excludes the possibility of the transformation into an antiferromagnetically modified state as mentioned above and provides evidence for the robustness of ferromagnetism in this system.
Here, we should note that the homogeneous ferromagnetism over the sample volume is evidenced from the NQR experiment using our sample that shows the absence of a paramagnetic signal below 1~K~\cite{Ohta2010}.

In Fig.~\ref{fig:equilibrium}(a), we show the quantity $M_{\rm FM+SC}-M_{\rm FM}$ that can be a measure of the superconducting contribution to the magnetization~\cite{Tachiki1979}, where $M_{\rm FM+SC}$ and $M_{\rm FM}$ denote the magnetization of the superconducting state and the normal state (at $T = 0.6$~K), respectively. 
We find that irreversibility develops, as observed in a mixed state of a conventional superconductor~\cite{Bean1964}. 
Remembering the result of Fig.~\ref{fig:M-H}(b), the development becomes evident in the superconducting state. 
Because of the absence of $H_{\rm c1}$, this is ascribed to the pinning and trapping effects of the vortex due to the spontaneous magnetization.

Next, we focus on the equilibrium magnetization $M_{\rm eq}(H)$ obtained by averaging the increasing- and decreasing-field curves.
In Fig.~\ref{fig:equilibrium}(b), we show $M_{\rm eq}(H)$ at $T = 0.07$~K ($< T_{\rm SC}$) and $0.6$~K ($> T_{\rm SC}$). 
We also plot $(M_{\rm FM+SC}-M_{\rm FM})_{\rm eq}$ at $T = 0.07$~K in Fig.~\ref{fig:equilibrium}(c). 
At a first glance, we note that there is no Meissner region in $M_{\rm eq}(H)$.
This supports the absence of $H_{c1}$ suggested from the ac susceptibility measurement. 
To firmly confirm this, we consider the magnetization curves $M_{\rm eq}(H)$ and $(M_{\rm FM+SC}-M_{\rm FM})_{\rm eq}$ expected for the cases of the presence and absence of $H_{c1}$, which are hereafter denoted case I and case II, respectively~\cite{Sonin1998}.    
For case I (see Figs.~\ref{fig:equilibrium}(d) and \ref{fig:equilibrium}(e)), reflecting the presence of the Meissner state ($B=H+4\pi M=0$), there is a linear region of the magnetization ($M=-1/4\pi H$) at low fields between $-H_{c1}$ and $H_{c1}$. 
This clearly contradicts the experiment. 
In contrast, the experiment is consistent with Figs.~\ref{fig:equilibrium}(f) and \ref{fig:equilibrium}(g), i.e., case II. 
One may note that the magnetization jump expected at zero field is not observed in the experimental result shown in Fig.~\ref{fig:equilibrium}(c). This is safely attributed to the field distribution of about 30~Oe arising from the field gradient of the experiment (see above).
As a consequence, the analysis of the equilibrium magnetization supports the observation of the ferromagnetism at $H \approx 0$ by the ac susceptibility measurements. 
We conclude that the lower critical field $H_{c1}$ is absent and the vortices are spontaneously created without external magnetic fields.

The spontaneous vortex state emerges when the condition $H_{c1}^{0} < 4\pi M_{\rm FM}$ is satisfied, where $H_{c1}^{0}$ is a lower critical field in a hypothetical nonmagnetic superconducting state~\cite{Tachiki1980,Sonin1998}. 
Note that $4\pi M_{\rm FM}$ plays a role of external field ($H_{c1} \simeq H_{c1}^{0} - 4\pi M_{\rm FM}$) and is estimated to be approximately 90~Oe from Fig.~\ref{fig:Arrott}(b). 
Therefore, $H_{c1}^{0}(0)$ should be greater than 90~Oe if it were present. 
However, $H_{c1}^{0}(0)$ seems to be at most on the order of several Oe as evaluated from a thermodynamic critical field $H_{c}(0) \sim 74$~Oe (that was calculated from our specific heat data of sample \#2 by assuming vacuum permeability) and a Ginzburg-Landau parameter $\kappa(0) \sim 48$ (that was calculated from $H_{c2}^{0}(0) \simeq H_{c2}(0) \sim 5$~kOe for the $c$-axis~\cite{Huy2008}). 
Consequently, the lower critical field cannot be present in the system.

The self-induced vortex state without the Meissner state seems consistent with the existence of two components below $T_{\rm SC}$ in the $1/T_1$ of the NQR experiment~\cite{Ohta2010}.
The observation of a slight increase in muon decay rate below $T_{\rm SC}$ might also be related to the distribution of internal fields created by the self-induced vortex state~\cite{Visser2009}. 
Nevertheless, we do not exclude other possibilities that have not been explored yet.
We hope that the present result stimulates further theoretical investigations of the coexistence of superconductivity and ferromagnetism, which may be closely related to the superconducting mechanism of UCoGe.

In conclusion, we reported the ac magnetic susceptibility and dc magnetization measurements of the superconducting ferromagnet UCoGe at low temperatures. 
In particular, we stress that this is the first measurement of dc magnetization in the superconducting state, to our knowledge. 
Both of these experiments show the absence of the Meissner state, equivalently the absence of the superconducting lower critical field $H_{c1}$. 
We also found that the ferromagnetic order robustly survives in the superconducting state, confirming the intrinsic coexistence of superconductivity with ferromagnetism. 
Furthermore, we observed that the hysteresis loop of the magnetization curve becomes strikingly enhanced in the superconducting state. 
These results are all compatible with the self-induced vortex state without the Meissner state. 
As a result, the novel state, which was theoretically proposed three decades ago but not experimentally established yet, is now found to be realized in UCoGe. 
Further work will be important to detect the quantized magnetic induction in the self-induced vortex state.

The authors thank K. Ishida, T. Ohta, H. Hattori, and K. Tenya for valuable discussions. 
This work was financially supported by Grants-in-Aid for Scientific Research (S) (No. 20224015) and for Scientific Research on Innovative Areas ``Heavy Electrons" (No. 20102006). 
This work was performed under the Inter-university Cooperative Research Program of the Institute for Materials Research, Tohoku University.
KD was supported by a Grant-in-Aid for Young Scientists (B) (No. 19740206) from MEXT.

\bibliography{16154} 


\end{document}